\documentclass[preprint,tightenlines,aps,floats,nofootinbib,showpacs,12pt]{revtex4}
\pdfoutput=1

\usepackage{graphicx}
\usepackage{amssymb}
\usepackage{slashed}

\newcommand{\isajet}{\texttt{ISAJET}}

\newcommand{\glui}{\tilde{g}}
\newcommand{\neut}[1]{\tilde{Z}_{#1}}
\newcommand{\charg}[1]{\tilde{W}_{#1}}

\newcommand{\meff}{M_{\textrm{eff}}}
\newcommand{\gev}{\textrm{GeV}}
\newcommand{\tg}{\tilde{g}}
\def\eslt{\slashed{E}_T}
\def\to{\rightarrow}

\def\bi{\begin{itemize}}
\def\ei{\end{itemize}}

\def\sps1ap{SPS1a$^\prime$}
\def\c1p{C1$^\prime$}

\def\tst{\tilde t}

\def\tg{\tilde g}

\def\tq{\tilde q}
\def\tw{\widetilde W}
\def\tz{\widetilde Z}
\def\thetab{\bar{\theta}}
\def\CM{{\cal M}}
\def\alt{\lesssim}
\def\agt{\gtrsim}
\def\be{\begin{equation}}  
\def\ee{\end{equation}}  
\def\bea{\begin{eqnarray}}
\def\eea{\end{eqnarray}}

\begin{document}

\title{Distinguishing LSP archetypes \\
via gluino pair production at LHC13}

\author{
Baris Altunkaynak$^{a}$\footnote{Email: baris@nhn.ou.edu},
Howard Baer$^{a}$\footnote{Email: baer@nhn.ou.edu},
Vernon Barger$^{b}$\footnote{Email: barger@pheno.wisc.edu},
Peisi Huang$^{c,d}$\footnote{Email: peisi@uchicago.edu}
}

\affiliation{
$^a$Homer L. Dodge Department of Physics and Astronomy,
University of Oklahoma, Norman, OK 73019, USA \\
$^b$Dept. of Physics, University of Wisconsin, Madison, WI 53706, USA \\
$^c$The Enrico Fermi Institute, University of Chicago, Chicago, IL 60637, USA \\
$^d$HEP Division, Argonne National Laboratory, Argonne, IL 60439, USA
}

\date{\today}

\begin{abstract}
The search for supersymmetry at Run 1 of LHC has resulted in gluino mass limits
$m_{\tg}\agt 1.3$ TeV for the case where $m_{\tq}\gg m_{\tg}$ and in models with
gaugino mass unification. The increased energy and ultimately luminosity of
LHC13 will explore the range $m_{\tg}\sim 1.3-2$ TeV. We examine how the
discovery of SUSY via gluino pair production would unfold via a comparative
analysis of three LSP archetype scenarios:
1. mSUGRA/CMSSM model with a bino-like LSP, 
2. charged SUSY breaking (CSB) with a wino-like LSP and 
3. SUSY with radiatively-driven naturalness (RNS) and a higgsino-like LSP. 
In all three cases we expect heavy-to-very-heavy squarks as suggested by a
decoupling solution to the SUSY flavor and CP problems and by the gravitino
problem. For all cases, initial SUSY discovery would likely occur in the
multi-$b$-jet + $\eslt$ channel. The CSB scenario would be
revealed by the presence of highly-ionizing, terminating tracks from
quasi-stable charginos. As further data accrue, 
the RNS scenario with 100-200 GeV higgsino-like LSPs
would be revealed by the build-up of a mass edge/bump in the OS/SF dilepton
invariant mass which is bounded by the neutralino mass difference. The
mSUGRA/CMSSM archetype would contain neither of these features but would be
revealed by a buildup of the usual multi-lepton cascade decay signatures.
\end{abstract}

\preprint{CETUP2015-005}


\maketitle

\section{Introduction}
\label{sec:intro}

The LHC8 (LHC with $\sqrt{s}=7-8$ TeV) era has come to a close and the LHC13 era
is underway! What have we learned from LHC8? The Standard Model (SM) has been
spectacularly confirmed in a vast assortment of measurements\cite{hinchliffe}.
And most importantly, a very SM-like Higgs boson has been revealed with mass
$m_h=125.09\pm 0.24$ GeV (ATLAS/CMS combined)\cite{atlas_h,cms_h}. The next
major target for LHC is to root out evidence for supersymmetry (SUSY). Indeed,
it has been declared that if LHC13 does not uncover evidence for SUSY early
within Run 2, then physics will have entered a state of
crisis\cite{Lykken:2014bca}!

What we have learned from LHC8 is that-- in generic models such as mSUGRA/CMSSM-- 
no evidence for SUSY translates into mass bounds of 
\bea
m_{\tg}\agt 1.3\ {\rm TeV} &&\ \ \ {\rm for}\ m_{\tq}\gg m_{\tg}\ \ \ {\rm and}\\
m_{\tg}\agt 1.8\ {\rm TeV} && \ \ \ {\rm for}\ m_{\tq}\sim m_{\tg} .
\eea
In addition, the rather large value of $m_h\simeq 125$ GeV seems to require
large radiative corrections to $m_h^2$ in the MSSM\cite{mhiggs}. The Higgs mass
can be accommodated with TeV-scale top squarks for large trilinear soft breaking
parameter $A_0$\cite{h125}, or by 10-100 TeV top squarks in the minimal mixing
case\cite{Arbey:2011ab}. Naively, these rather high sparticle mass limits seem
to conflict with notions of weak scale naturalness which favor sparticles at or
around $m_{weak}\simeq 100$ GeV, the value of $m_{W,Z,h}$. This has led to some
puzzlement as to the emerging Little Hierarchy: why is $m(sparticle)\gg
m_{weak}$? It has also led to more detailed examination of what is meant by
electroweak naturalness.

The point of contact between SUSY Lagrangian mass parameters (soft terms and
superpotential $\mu$ term) and hard data occurs in the scalar (Higgs) potential:
in the MSSM, it is given by
\be
V_{Higgs}=V_{tree}+\Delta V,
\ee
where the tree level portion is given by
\bea
V_{tree} &= &(m_{H_u}^2+\mu^2)|h_u^0|^2 +(m_{H_d}^2+\mu^2)|h_d^0|^2 \nonumber \\ 
&& -B\mu (h_u^0h_d^0+h.c.)+{1\over 8}(g^2+g^{\prime 2})
(|h_u^0|^2-|h_d^0|^2)^2
\eea
and the radiative corrections (in the effective potential approximation) by
\be
\Delta V=\sum_{i}\frac{(-1)^{2s_i}}{64\pi^2}Tr\left((\CM_i\CM_i^\dagger)^2
\left[\log {\CM_i\CM_i^\dagger\over Q^2}-{3\over 2}\right]\right) ,
\ee
where the sum over $i$ runs over all fields that couple to Higgs fields,
$\CM_i^2$ is the {\it Higgs field dependent} mass squared matrix (defined as the
second derivative of the tree level Lagrangian) of each of these fields, and the
trace is over the internal as well as any spin indices. Minimization of the
scalar potential in the $h_u^0$ and $h_d^0$ directions allows one to compute the
gauge boson masses in terms of the Higgs field vacuum expectation values $v_u$
and $v_d$, and leads to the well-known condition that
\be
\frac{m_Z^2}{2} = \frac{(m_{H_d}^2+\Sigma_d^d)-(m_{H_u}^2+\Sigma_u^u)\tan^2\beta}{(\tan^2\beta -1)}
-\mu^2 
\label{eq:mzs},
\ee 
where the $\Sigma_u^u$ and $\Sigma_d^d$ terms arise from derivatives of $\Delta
V$ evaluated at the potential minimum and $\tan\beta\equiv\frac{v_u}{v_d}$. This
minimization condition relates the $Z$-boson mass scale to the soft SUSY
breaking terms and the superpotential higgsino mass $\mu$. In most computations
of the SUSY mass spectrum, the weak scale soft terms are determined by
renormalization group running from a constrained set of parameters set at some
high scale $\Lambda$. In gravity-mediation\cite{sugra}, $\Lambda$ is usually
taken to be $m_{GUT}\simeq 2\times 10^{16}$ GeV\cite{sugreview}. Then the weak
scale value of $\mu$ is dialed (fine-tuned) so that the measured value of $m_Z$
is obtained. An evaluation of the extent of this fine-tuning is provided by the
{\it electroweak} measure $\Delta_{EW}$ which evaluates the largest of the 43
terms on the right-hand-side of Eq. \ref{eq:mzs}. 
If one term on the RHS is $\gg m_Z^2$, then some other unrelated term will have
to be large and of opposite-sign to guarantee that $m_Z=91.2$ GeV. To avoid such
large weak scale tuning, evidently all terms on the right-hand-side of Eq.
\ref{eq:mzs} should be comparable to or less than $m_Z^2$. This implies the
following\cite{ltr,rns}.
\bi
\item The superpotential higgsino mass $\mu\sim 100-200$ GeV, the closer to
$m_Z$ the better. The lower limit $\mu\agt 100$ GeV comes from null searches for
chargino pair production at LEP2.
\item The soft term $m_{H_u}^2$ is radiatively driven to small values $\sim
-m_Z^2$ at the weak scale.
\item The radiative corrections $\Sigma_u^u$ are not too large. The largest of
these usually comes from the top-squarks. Each of the terms
$\Sigma_u^u(\tst_{1,2})$ are minimized by TeV-scale highly mixed top squarks, a
condition which also lifts $m_h$ up to $\sim 125$ GeV\cite{ltr,rns}.
\ei

Some alternative fine-tuning measures also have been advocated in the literature. 
\begin{enumerate}
\item The usual application of the Higgs mass large-log measure
$\Delta_{HS}=\delta m_{H_u}^2/(m_h^2/2)$ where $\delta m_{H_u}^2\sim
\frac{f_t^2}{8\pi^2}\left( m_{Q_3}^2+m_{U_3}^2+A_t^2\right) \ln \left( \Lambda
/m_{SUSY}\right)$ has been challenged\cite{comp3,seige} in that it ignores the
{\it dependent} term $m_{H_u}^2$ which occurs in the RGE. But the larger
$m_{H_u}^2(\Lambda )$ becomes, the greater is the cancelling correction to
$\delta m_{H_u}^2$\cite{arno}. By appropriately combining dependent terms,
$\Delta_{HS}$ reduces to the same general consequences as $\Delta_{EW}$. 
\item Alternatively, the Ellis {\it et al.}/Barbieri-Giudice
measure\cite{eenz,bg} is defined as $\Delta_{BG}\equiv \max_i|\partial \ln (m_Z^2
)/\partial p_i |$ where the $p_i$ constitute fundamental high scale parameters
of the theory. To evaluate $\Delta_{BG}$, $m_Z^2$ must be evaluated in terms of
fundamental high scale parameters usually taken to be the GUT scale soft
breaking terms. The usual application of this measure has been
challenged\cite{comp3,seige} in that in supergravity theories, the soft terms
are {\it not independent}, but are evaluated as multiples of the fundamental
gravitino mass $m_{3/2}$. Evaluating $\Delta_{BG}$ in terms of the {\it
independent} parameters $\mu$ and $m_{3/2}$, low $\Delta_{BG}$ also leads to the
same general consequences as $\Delta_{EW}$.
\end{enumerate}

Using $\Delta_{EW}$, then indeed most constrained high scale SUSY models are
found to be highly tuned in the EW sector\cite{seige}. An exception occurs for a
pocket of parameter space of the two-extra parameter non-universal Higgs
models\cite{nuhm2} where $\mu\sim 100-200$ GeV and where $m_{H_u}^2$ is driven
to small negative values comparable to $-m_Z^2$ while allowing for highly mixed
TeV-scale top squarks which provide $m_h\simeq 125$ GeV. This pocket of
parameter space we call SUSY with radiatively-driven naturalness, or RNS for
short. By requiring EW naturalness, then upper bounds can be computed for all
sparticle masses\cite{rns}. In radiative natural SUSY with $\Delta_{EW}<10\
(30)$ then it is found that\cite{rns}:
\bi
\item $m_{\tg}\alt 2.5\ (5)$ TeV,
\item $m_{\tst_1}\alt 2\ (3)$ TeV,
\item $m_{\tw_1,\tz_{1,2}}\alt 200\ (300)$ GeV.
\ei
The first of these values can be compared to the ultimate reach of LHC14 with
1000 fb$^{-1}$ where a $5\sigma$ discovery can be established for $m_{\tg}\alt
2$ TeV\cite{bblt,rns@lhc}. Thus, while EW naturalness certainly allows for
gluinos and squarks to lie well beyond the ultimate reach of LHC13, it is also
true that the {\it most} natural values of gluino and squark masses are those
within the exploratory range of LHC13: the lighter the better. This motivates an
examination of how a SUSY discovery via gluino pair production is likely to
unfold at LHC13 when the gluino mass lies just beyond present bounds.

In this paper we assume a gluino mass of $m_{\tg}=1400$ GeV, {\it i.e.} just
beyond present bounds. We then investigate how a SUSY discovery would unfold
under three lightest SUSY particle (LSP) archetype scenarios:
\bi
\item the mSUGRA/CMSSM model with a bino-like LSP,
\item a charged SUSY breaking (CSB) scenario with a wino-like LSP and
\item SUSY with radiatively-driven naturalness and a higgsino-like LSP.
\ei
Our goal is to look for commonalities and differences between these three
archetype scenarios that would allow a rapid determination of the nature of the
LSP if a gluino pair production signal emerges at LHC13.

Towards this end, in Sec. \ref{sec:BM}, we present three archetype benchmark
models (BM) labelled as mSUGRA, CSB and RNS. While each BM model contains a gluino
with mass 1400 GeV, their implications for collider searches will be very
different. In Sec. \ref{sec:results}, we discuss how SUSY discovery would unfold
in each BM model while in Sec. \ref{sec:LSP} we discuss how each archetype could
ultimately be distinguished as more integrated luminosity accrues. Briefly, in
all cases the most likely initial discovery channel could occur in the
multi-$b$-jet + $\eslt$ channel with $\sim 3-8$ fb$^{-1}$ of integrated
luminosity. For the CSB benchmark, the model would be distinguished by the
presence of one or more {\it cm}-length highly ionizing tracks (HITs) from
quasi-stable charginos which are produced within the gluino cascade decays. For
the RNS scenario, as 100-1000 fb$^{-1}$ of integrated luminosity accumulates, then
a distinctive opposite-sign/same flavor (OS/SF) dilepton invariant mass edge
should develop in multi-$b$-jet + $\eslt$ events which contain such a dilepton
pair. The mass edge occurs at the kinematic limit $m(\ell^+\ell^-
)<m_{\tz_2}-m_{\tz_1}\sim 10-30$ GeV in RNS models. For the mSUGRA benchmark,
neither of the above distinctive features should develop, but instead the usual
multi-lepton plus multijet + $\eslt$ cascade decay topologies should build up as
greater integrated luminosity accrues. Our summary and conclusions are given in
Sec. \ref{sec:conclude}.

\section{Benchmark Models}
\label{sec:BM}

In this section, we present three benchmark models representing each of three
LSP archetype scenarios. Each scenario contains a light Higgs scalar $m_h\simeq
125$ GeV\footnote{We allow for a $\sim \pm 2$ GeV theory uncertainty on the
Isajet RG-improved one loop effective potential calculation of $m_h$ which
includes leading two-loop terms\cite{mhiggs}.} and a gluino of mass
$m_{\tg}=1.4$ TeV, just beyond the bounds from LHC8. All spectra were generated
using the Isajet/Isasugra 7.84 program\cite{isajet}.

\subsection{mSUGRA/CMSSM}

In the minimal supergravity model (mSUGRA or CMSSM)\cite{sugra}, it is assumed
that supergravity is broken in a hidden sector leading to a massive gravitino
characterized by mass $m_{3/2}$, with $m_{3/2}\sim 1$ TeV in accord with
phenomenological requirements. In the limit as $M_P\to\infty$ but keeping
$m_{3/2}$ fixed, then one is lead to the global SUSY Lagrangian of the MSSM
augmented by soft SUSY breaking terms each of order $m_{3/2}$. A simplifying
assumption (with minimal motivation) is that all soft scalar masses are unified
to $m_0$ at the GUT scale. In addition, all gaugino masses are unified to
$m_{1/2}$, all trilinears are unified to $A_0$ and there is a bilinear term $B$.
Renormalization group running connects the GUT scale parameters to the weak
scale ones. At the weak scale, the scalar potential is minimized and the
superpotential $\mu$ parameter is dialed (fine-tuned) so as to generate the
measured value of $m_Z=91.2$ GeV.

\begin{figure}[!htb]
\begin{center}
\includegraphics[width=160mm]{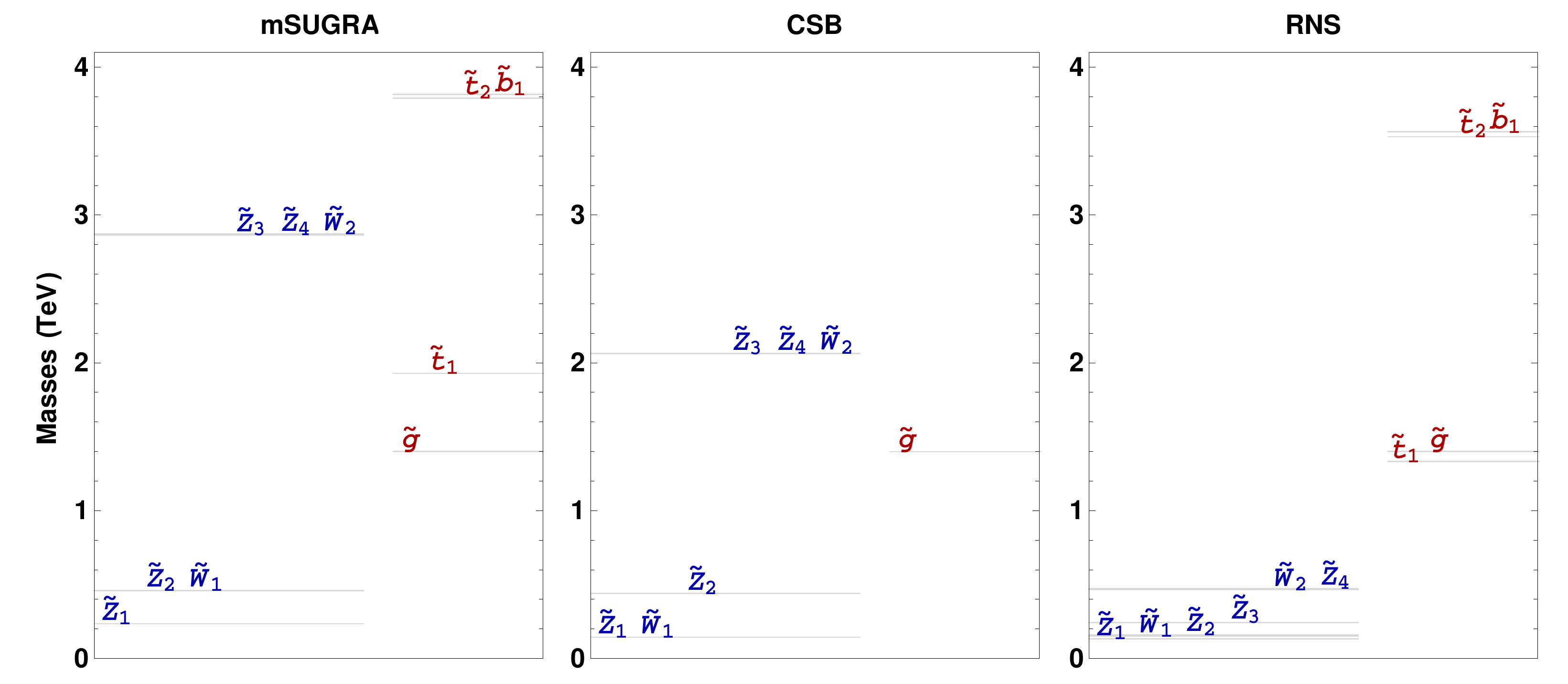}
\caption{Spectra of benchmark models: mSUGRA (left), CSB (middle), RNS (right).
We display the bottom part of the spectra up to 4 TeV where the electroweak
-ino's are shown in blue while the gluinos, stops and sbottoms are shown in red.
The sparticles within the same column are ordered in increasing mass from left
to right. In the CSB model, the stop and sbottom masses are $\sim 35$ TeV: see
Table~\ref{tab:BM}.}
\label{fig:BM}
\end{center}
\end{figure}

Spectra from this popular
model~\cite{Ross:1992tz,Arnowitt:1992aq,Barger:1992ac,Barger:1993gh} can be
generated with many computer codes. In Table \ref{tab:BM}, we show a mSUGRA
benchmark model with $m_0=5$ TeV, $m_{1/2}=517$ GeV, $A_0=-8$ TeV and the ratio
of Higgs vevs $\tan\beta =10$. These parameters lead to a spectra with a gluino
mass $m_{\tg}=1.4$ TeV, {\it i.e.} just beyond the reach of LHC8. The light
Higgs mass $m_h=123.6$ GeV, in accord with its measured value if one allows for
the $\pm 2$ GeV uncertainty in our calculation of $m_h$. The $\tz_1$ is a
bino-like LSP. The superpotential $\mu$ parameter turns out to be $\mu =2861$
GeV leading to $\Delta_{EW}= 1968$ so that this benchmark is highly fine-tuned
in the EW sector. The calculated thermal neutralino abundance
$\Omega_{\tz_1}^{TP}h^2=317$, far beyond the measured value.
Thus, some sort of 
1. late entropy dilution, 
2. decay of $\tz_1$ to an even lighter LSP such as an axino or 
3. $R$-parity violating decays of $\tz_1$ would need to be invoked to bring the
model into accord with the measured dark matter density. 
A schematic illustration of the lighter spectral states of the mSUGRA benchmark
is shown in Fig. \ref{fig:BM}.

\begin{table}[!htb]
\renewcommand{\arraystretch}{1.2}
\begin{center}
\begin{tabular}{c|ccc}
 & mSUGRA & CSB & RNS \\
\hline \hline
$m_0$ & 5,000 & 50,570 & 5,000 \\
$M_1$ & 517.0 & 927.3 & 517.8 \\
$M_2$ & 517.0 & 140.5 & 517.8 \\
$M_3$ & 517.0 & -421.5 & 517.8 \\
$A_0$ & -8,000 & 140.5 & -8,000 \\
$\tan\beta$ & 10 & 10 & 10 \\
\hline
$\mu$ & 2,861 & 2,000  & 150 \\
$m_A$ & 5,666 & 2,000  & 2,000 \\
$m_h$ & 123.6 & 126.4  & 124.1 \\
$m_{\glui}$ & 1,400 & 1,399 & 1,399 \\
$m_{\tilde{u}_L}$ & 5,065 & 50,205 & 5,038 \\
$m_{\tilde{t}_1}$ & 1,929 & 34,327 & 1,332 \\
$m_{\charg2}$ & 2.872.0 & 2,064.8 & 464.3 \\
$m_{\charg1}$ & 460.8 & 143.6 & 150.7 \\
$m_{\neut4}$ & 2,866.3 & 2,062.8 & 473.6 \\
$m_{\neut3}$ & 2,865.1 & 2,062.2 & 243.3 \\
$m_{\neut2}$ & 459.8 & 438.8 & 159.5 \\
$m_{\neut1}$ & 234.3 & 143.4 & 132.1 \\
\hline
Bino frac.     & 0.9999 & 0.0022 & 0.2915 \\
Wino frac.     & 0.0010 & 0.9993 & 0.1747 \\
Higgsino frac. & 0.0151 & 0.0365 & 0.9405 \\
\hline
$\Omega_{\tz_1}^{TP}h^2$ & 317 & 0.0013 & 0.01 \\
$\Delta_{EW}$ & 1968 & 5228 & 10.4 \\
\hline
\end{tabular}
\caption{Input parameters and masses (in GeV) for three benchmark points
computed with \isajet{} 7.84 \cite{isajet}. Also displayed are the bino, wino
and Higgsino fractions.}
\label{tab:BM}
\end{center}
\end{table}

\subsection{Charged SUSY breaking}

In models labeled as minimal anomaly-mediation (mAMSB)\cite{amsb}, it is assumed
that SUSY is broken in a secluded sector so that the dominant contributions to
soft terms come not from tree-level supergravity but from the superconformal
anomaly. Such models leads characteristically to spectra including wino-like
gauginos as the lightest SUSY particles\cite{amsb_pheno}. Further, one obtains
spectra with well-known tachyonic sleptons. In the original
construct\cite{amsb}, it was suggested to augment soft scalar masses with a
common $m_0^2$ term to cure the tachyon problem.

The original mAMSB models seem disfavored in that they have problems generating
$m_h\simeq 125$ GeV due to a rather small weak scale $A_t$ soft
term\cite{Arbey:2011ab,dm125}. An alternative incarnation goes under the label
of PeV SUSY\cite{wells}, split SUSY\cite{split}, pure gravity
mediation\cite{yanagida} and spread SUSY\cite{hall}. In the simple yet elegant
construction of Wells\cite{wells}, it is argued that the PeV scale (with $m({\rm
scalars})\sim m_{3/2}\sim 1$ PeV$=$1000 TeV) is motivated by considerations of
wino dark matter and neutrino mass while providing a decoupling
solution~\cite{dine} to the SUSY flavor, CP, proton decay and gravitino/moduli
problems. This model invoked ``charged SUSY breaking'' (CSB) where the hidden
sector superfield $S$ is charged under some unspecified symmetry. In such a
case, the scalars gain masses via SUGRA 
\be
\int d^2\theta d^2\thetab \frac{S^\dagger S}{M_P^2}\Phi_i^\dagger \Phi_i
\Rightarrow
\frac{F_S^\dagger F_S}{M_P^2}\phi_i^*\phi_i
\ee
while gaugino masses, usually obtained via gravity-mediation as
\be
\int d^2\theta\frac{S}{M_P}WW\Rightarrow \frac{F_s}{M_P}\lambda\lambda ,
\ee
are now forbidden. Then the dominant contribution to gaugino masses comes from
AMSB:
\bea
M_1&=& \frac{33}{5}\frac{g_1^2}{16\pi^2}m_{3/2}\sim m_{3/2}/120,\\
M_2&=& \frac{g_2^2}{16\pi^2}m_{3/2}\sim m_{3/2}/360,\\
M_3&=& -3\frac{g_3^2}{16\pi^2}m_{3/2}\sim -m_{3/2}/40 .
\eea
Saturating the measured dark matter abundance with thermally-produced (TP) winos
requires $m_{\tw}\sim M_2 \sim 2.5$ TeV which in turn requires the gravitino and
scalar masses to occur at the $\sim 1000$ TeV (1 PeV) level. A virtue of the CSB
model is that the highly massive top squarks $m_{\tst_{1,2}}\sim 50-100$ TeV
lead to $m_h\sim 125$ GeV even with a tiny $A_t$ trilinear soft term.

The CSB benchmark point is listed in Table \ref{tab:BM} where $m_0\simeq
m_{3/2}=50.57$ TeV leading to squark and slepton masses $\sim 50$ TeV but with
$m_{\tg}=1.4$ TeV. The LSP is a wino-like neutralino $\tz_1$ with mass
$m_{\tz_1}=143.4$ GeV. The superpotential $\mu$ parameter is taken to be $2$
TeV. The dominant contribution to the EW fine-tuning measure $\Delta_{EW}$ comes
from the top squark radiative corrections leading to $\Delta_{EW}=5228$ so the
model is highly fine-tuned in the EW sector. The thermally produced wino-like
neutralino abundance is found from IsaReD\cite{isared} to be
$\Omega_{\tz_1}^{TP}=0.0013$ so WIMPs are thermally underproduced. They could be
augmented via non-thermal WIMP production ({\it e.g.} from gravitino, axino,
saxion or moduli decays\cite{moroi_randall}) or the DM abundance could be
augmented by other species such as axions\cite{winoaxion}. The CSB benchmark is
also shown schematically in Fig. \ref{fig:BM}.

\subsection{SUSY with radiatively-driven naturalness (RNS)}

In models with radiatively-driven naturalness, it is assumed that soft terms
arise via gravity mediation and are characterized by the scale $m_{3/2}\sim
2-20$ TeV. Such heavy soft terms lead to $m_h\simeq 125$ GeV for highly mixed
TeV-scale top squarks. The $\mu$ parameter arises differently. In the SUSY DFSZ
axion model\cite{dfsz,susydfsz}, the Higgs multiplets $\hat{H}_u$ and
$\hat{H}_d$ are assigned PQ charges so that the usual $\mu$ term is forbidden
although now the Higgs superfields may couple to additional gauge singlets from
the PQ sector. The $\mu$ term is then re-generated via $PQ$ symmetry breaking at
a value of $\mu\sim f_a^2/M_P$ so that the Little Hierarchy $\mu\ll m_{3/2}$ is
merely a reflection of the mis-match between PQ breaking scale and hidden sector
mass scale $f_a\ll m_{hidden}$. In the MSY SUSY axion model\cite{msy}, the PQ
symmetry is broken radiatively as a consequence of SUSY breaking in a similar
manner that EW symmetry is radiatively broken as a consequence of SUSY breaking.
The radiative PQ breaking generates a small $\mu\sim 100-200$ GeV (as required
by naturalness) from multi-TeV values of $m_{3/2}$\cite{radpq}. Once $\mu$ is
known, then the weak scale value of $m_{H_u}^2$ is determined by the scalar
potential minimization condition and is also of order $-m_Z^2$ as required by
naturalness. The weak scale value of $m_{H_u}^2$ is evolved to $m_{GUT}$ where
it is found that $m_{H_u}(m_{GUT})\ne m_0$, where $m_0$ now labels just the
matter scalar masses. 

The RNS benchmark model is shown in Table \ref{tab:BM} with matter scalar mass
$m_0=5$ TeV and a trilinear soft term $A_0=-8$ TeV. The ratio of Higgs vevs
$\tan\beta =10$ and the pseudoscalar Higgs mass $m_A$ is taken as 2 TeV. The
unified gaugino mass $m_{1/2}=517.8$ GeV leading to $m_{\tg}=1.4$ TeV. The
highly mixed top squarks with mass $m_{\tst_{1,2}}=1.3 (3.5)$ TeV lead to
$m_h=124.1$ GeV. Since $\mu=150$ GeV, then the model has $\Delta_{EW}=10.4$ or
about 10\% fine-tuning in the EW sector: the model is very natural. The LSP is a
higgsino-like WIMP with mass $m_{\tz_1}=132.1$ GeV. The TP relic density
$\Omega_{\tz_1}^{TP}h^2=0.01$ but in this case the axion could comprise the bulk
of DM\cite{axdm}.\footnote{An alternative way to match the measured DM density
is to reduce the bino mass $M_1$ for the case of gaugino mass non-universality:
see \cite{Baer:2015tva}.} The RNS benchmark is schematically shown as the third
frame of Fig. \ref{fig:BM}.

\section{How SUSY discovery unfolds}
\label{sec:results}

\subsection{Gluino pair production}

In the benchmark scenarios we have selected, a heavy spectrum of matter
scalars-- squarks and sleptons-- is assumed. This is in accord with at least a
partial decoupling solution to the SUSY flavor, CP, gravitino and proton-decay
problems. In addition, to accommodate Affleck-Dine\cite{ad} leptogenesis, then a
non-flat K\"ahler metric is required\cite{drt} from which one would expect
generic flavor and CP violation. The decoupling solution allows the AD mechanism
to proceed in the face of potential flavor violations.

\begin{figure}[!htb]
\begin{center}
\includegraphics[width=110mm]{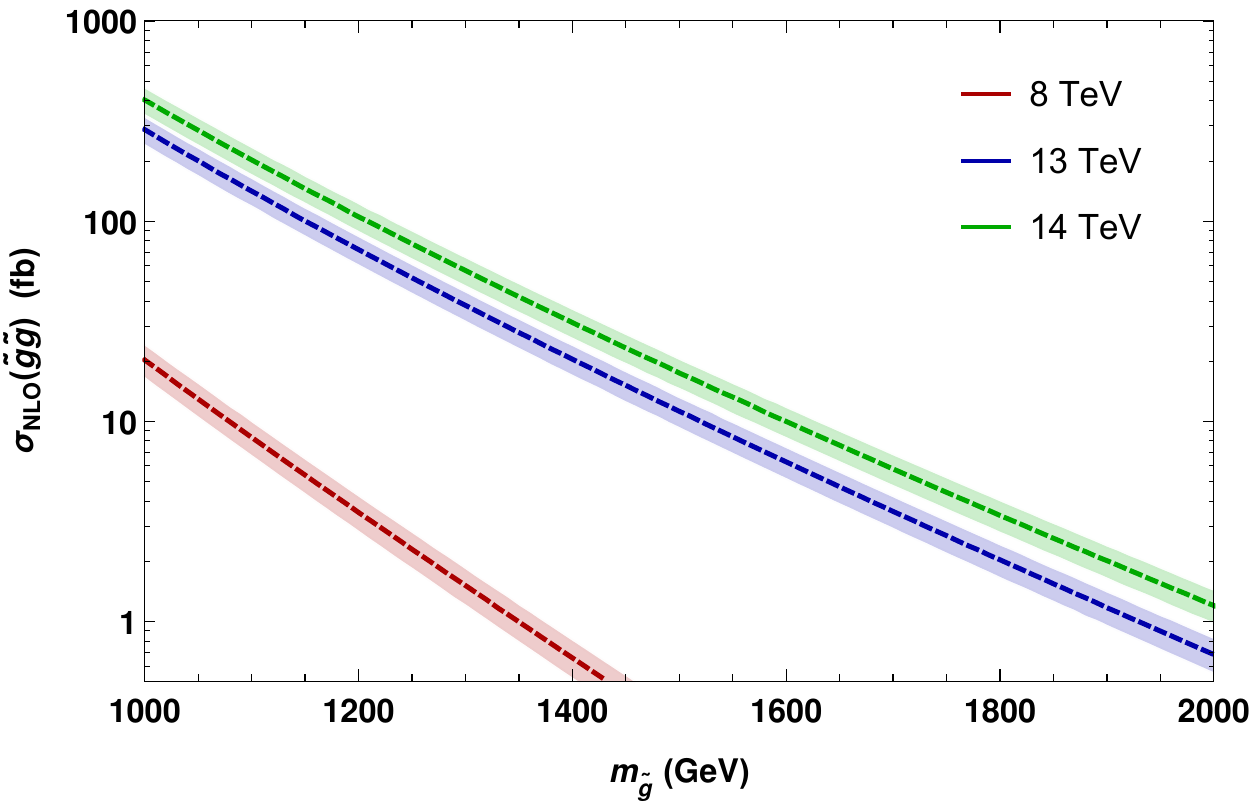}
\caption{Gluino pair production cross section at the LHC for $\sqrt{s} = 8, 13,
14$ TeV calculated at NLO with Prospino\cite{prospino}. Squarks are assumed to
be heavy with mass $m_{\tilde{q}} = 5$ TeV. The shaded areas show the scale
uncertainty.}
\label{fig:gg}
\end{center}
\end{figure}

In the case of decoupled matter scalars, then we expect gluino pair production
and possibly electroweak -ino pair production to offer the main SUSY discovery
reactions. In Fig. \ref{fig:gg}, we show the NLO values of $\sigma (pp\to \tg\tg
X)$ reaction versus $m_{\tg}$ for $\sqrt{s}=8$, 13 and 14 TeV. The squark masses
have been set to 5 TeV. We use Prospino to calculate the total cross
sections\cite{prospino}.

For our benchmark points with $m_{\tg}=1.4$ TeV, we see that the LHC8 total
production cross section $\sigma (\tg\tg )$ is about 0.6 fb. As $\sqrt{s}$ is
increased to 13 TeV for LHC Run 2, then the total gluino pair production cross
section jumps by a factor of $\sim 30$ to $\simeq 20$ fb. Future LHC runs with
fully trained magnets may attain $\sqrt{s}\sim 14$ TeV for which $\sigma (\tg\tg
)$ would rise to $\sim 35$ fb. While EW -ino pair production rates should be
comparable to gluino pair production-- due to their lower masses-- we expect at
this stage that gluino pair production is more easily seen due to its large
energy release and no cost for leptonic branching fractions in the major signal
channel of jets + $\eslt$.

\subsection{Gluino branching fractions and signatures}

Once produced, the gluinos can cascade decay\cite{cascade} to a variety of final
states which are listed in Table \ref{tab:BFs}. The decay modes including $q$ in
the final state are summed over $q=u,d,s,c$ possibilities. It is evident from
the Table that in all cases the decays to third generation quarks are enhanced
over first/second generation quarks. Gluino three-body decays to third
generation quarks were first calculated in Ref's \cite{btw,bartl,btw2} where
their enhancement was noticed to arise from 1. couplings which include the large
$b$ and $t$ Yukawa couplings, 2. generically smaller mediator masses
$m_{\tst_{1,2}}\alt m_{\tq}$ and 3. large L-R mixing effects. 
For our benchmark models, we see that in mSUGRA, the $\tg$ decays to states
including $b\bar{b}$ (both directly and via decay to top followed by $t\to bW$)
81\% of the time, while for CSB it is 47\% and for RNS it is 99.1\%. Thus, for
$\tg\tg$, we usually expect the presence of four $b$-jets in the final state
(although some of these may fall below acceptance cuts or be merged with other
$b$-jets {\it etc.}). In the CSB case, the branching to $t$ and $b$ quarks is
only mildly enhanced since all six squark flavors are extremely heavy. In
addition, in the mSUGRA and CSB cases, gluinos only decay substantially to the
lighter -ino states $\tw_1$ and $\tz_{1,2}$. For the RNS case, gluino decays to
the light higgsino-like EWinos dominates but also decays to the heavier bino-
and wino-like states $\tz_{3,4}$ and $\tw_2$ can be substantial.
\begin{table}[!htb]
\renewcommand{\arraystretch}{1.2}
\begin{center}
\begin{tabular}{c|ccc}
 final state & mSUGRA & CSB & RNS \\
\hline \hline
$q \bar{q}'\charg{1}$      & 10.5 \% & 34.0 \%  & 0.1 \% \\
$tb\charg{1}$              & 13.4 \% & 28.8 \%  & 45.6 \%\\
$tb\charg{2}$              & --   \% & --   \%  &  2.2 \%\\
$q \bar{q}\neut{1}$        & 3.1  \% & 17.0 \%  & --   \% \\
$b \bar{b}\neut{1}$        & 0.5  \% & 8.7  \%  & --   \% \\
$t \bar{t}\neut{1}$        & 60.3 \% & 6.2  \%  & 17.2 \% \\
$q \bar{q}\neut{2}$        & 5.2  \% & 2.4  \%  & --   \% \\
$b \bar{b}\neut{2}$        & 4.3  \% & 0.3  \%  & --   \% \\
$t \bar{t}\neut{2}$        & 2.5  \% & 3.0  \%  & 22.5 \% \\
$t \bar{t}\neut{3}$        & --   \% & --   \%  & 10.6 \% \\
$t \bar{t}\neut{4}$        & --   \% & --   \%  & 1.0  \% \\
\hline
\end{tabular}
\caption{Gluino branching fractions for the three benchmark models where $q =
u$, $d$, $c$ and $s$.}
\label{tab:BFs}
\end{center}
\end{table}

\begin{figure}[!htb]
\begin{center}
\includegraphics[width=100mm]{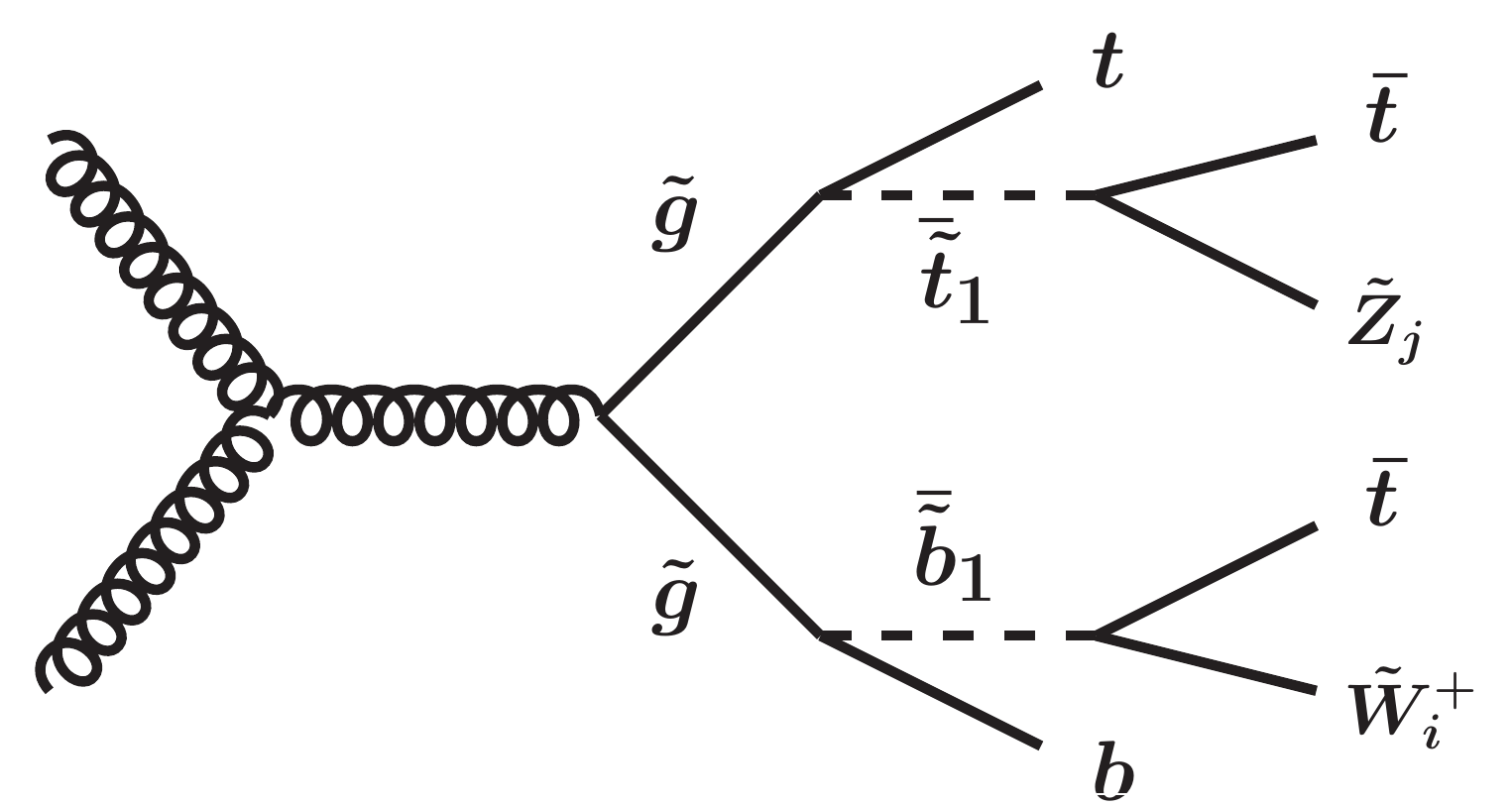}
\caption{Gluino pair production and decay to multiple $b$-jets in the three
benchmark scenarios.}
\label{fig:diagram}
\end{center}
\end{figure}

A diagram depicting gluino pair production followed by typical three-body decays
is shown in Fig. \ref{fig:diagram}. The presence of up to four $b$-jets in the
final state can be used as a powerful veto against dominant SM backgrounds such
as $t\bar{t}$ production. Indeed, ATLAS searches\cite{atlas_b} for $\tg\tg$
production with $\ge 3$ $b$-jets in the final state offers the most powerful
probe of gluino masses in the case where $m_{\tg}\ll m_{\tq}$.

\subsection{Gluino cascade decay signatures}

We use Isajet 7.84\cite{isajet} to generate a SUSY Les Houches Accord\cite{lha}
(SLHA) file for each benchmark scenario which is fed into Pythia\cite{pythia}
for generation of gluino pair production events followed by cascade decays. The
gluino pair cross section is normalized to the NLO Prospino results of Fig.
\ref{fig:gg}. We use the Snowmass SM background event
set\cite{Avetisyan:2013onh} for the background processes. The $t\bar{t}$
background set is expected to be the dominant background\cite{gabe}, where extra
$b$-jets can arise from initial/final state radiation and from jet mis-tags.
While the Snowmass background set was generated for $\sqrt{s}=14$ TeV LHC
collisions, we have re-scaled the rates for $\sqrt{s}=13$ TeV collisions. Our
signal and BG events are passed through the Delphes\cite{delphes} toy detector
simulation as set up for Snowmass analyses.

We apply the following event selection cuts:
\begin{itemize}
\item $n(\textit{jets}) \geq 4$ , 
\item $n(\textit{b-jets}) \geq 3$, 
\item $E_T(j_1, j_{2-4}) > 100, 50 \, \gev$,
\item for isolated leptons, then $p_T(\ell) > 20 \, \gev$,
\item $\eslt > \eslt(cut)=50,100-500 \, \gev$
\item $A_T > 1200 \, \gev$,
\end{itemize}
where $A_T = \eslt + \sum_{leptons} E_T + \sum_{jets} E_T$ and for later use
$\meff = \eslt + \sum_{i=1}^4 E_T(j_i)$. To gain some optimization of
signal-to-background (S/B), we tried the above range of $\eslt$ cuts and
evaluated S/B with and without the $A_T$ cut.

\begin{figure}[!htb]
\begin{center}
\includegraphics[width=110mm]{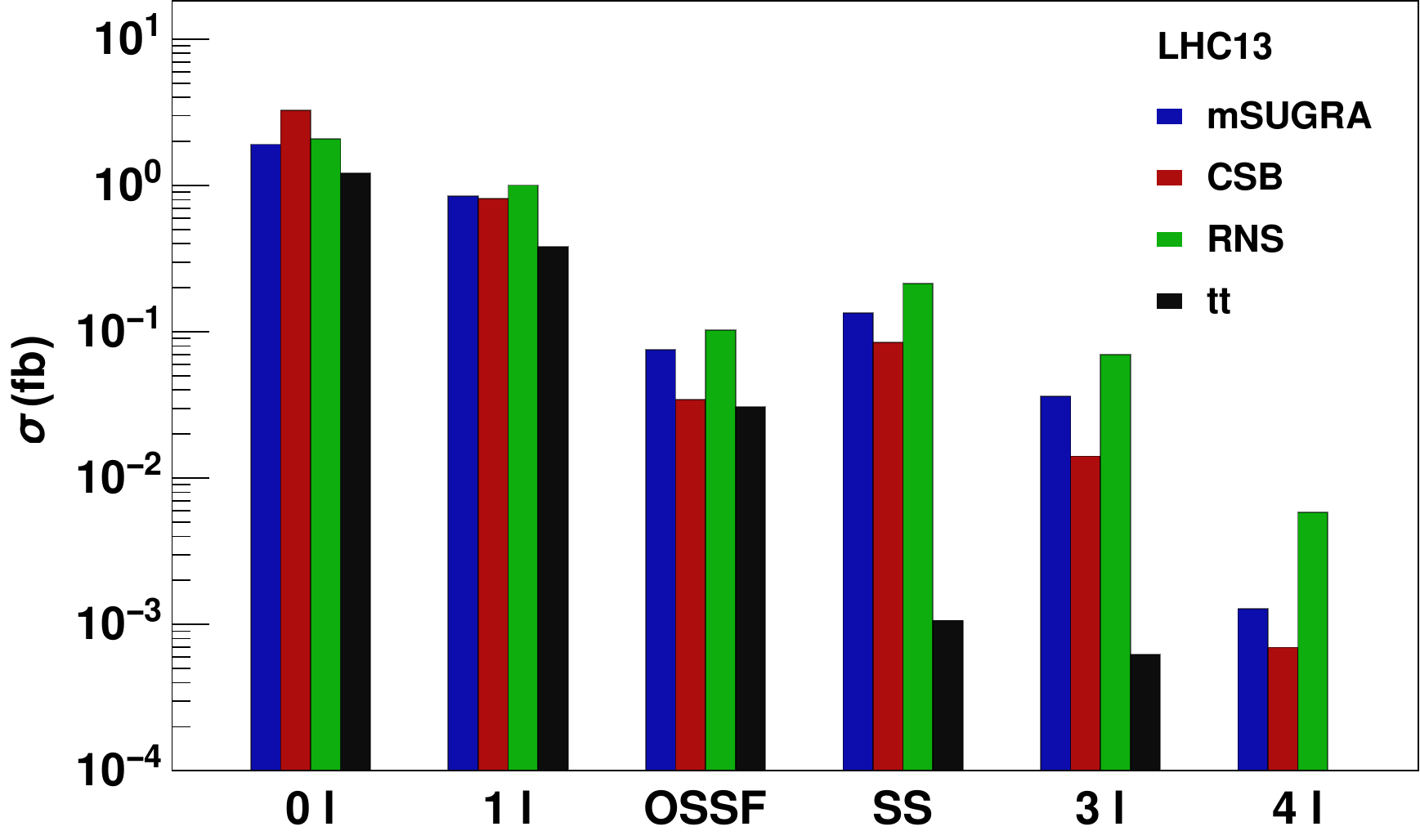}
\caption{Cross section after cuts from gluino pair production for the three SUSY
benchmark models and from $t\bar{t}$ background. For the $0$ and $1\ell$
signals, we take $\eslt >500$ GeV while for the OSSF dilepton channel we take
$\eslt > 400$ GeV. For the SS, $3\ell$ and $4\ell$ signals, we require $\eslt
>50$ GeV.}
\label{fig:sigma}
\end{center}
\end{figure}

The cross sections after cuts for various multi-lepton $+\ge 3$
$b$-jets + $\eslt$ channels are shown in Fig. \ref{fig:sigma}. The optimal
$\eslt$ cut for the $0\ell$ and $1\ell$ channels was the hardest value:
$\eslt >500$ GeV. For the Opposite Sign Same Flavor (OSSF) dilepton channel,
the best cut was $\eslt >400$ GeV while for the Same Sign ($SS$)-dilepton,
$3\ell$ and $4\ell$ channels, the $\eslt >50$ GeV was best. The $A_T>1200$ GeV
cut helped just marginally.

\begin{figure}[!htb]
\begin{center}
\includegraphics[width=110mm]{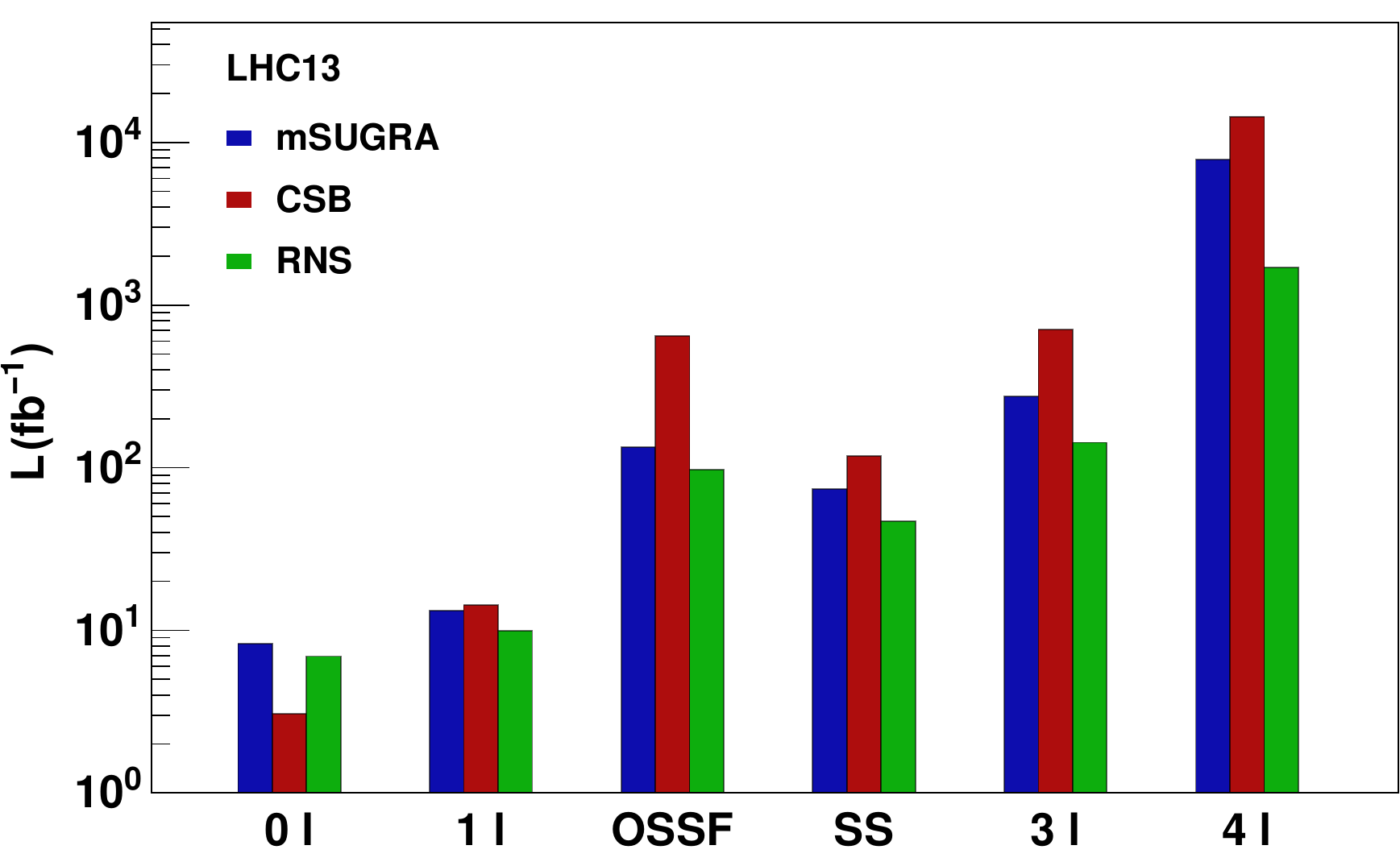}
\caption{Required integrated luminosity at LHC13 to establish a $5\sigma$ SUSY
discovery in various channels from gluino pair production for the three SUSY
benchmark models compared to $t\bar{t}$ background. For the $0$ and $1\ell$
signals, we take $\eslt >500$ GeV while for the OSSF dilepton channel we take
$\eslt > 400$ GeV. For the SS, $3\ell$ and $4\ell$ signals, we require $\eslt
>50$ GeV.}
\label{fig:L}
\end{center}
\end{figure}

We see, from Fig. \ref{fig:sigma}, that the signal cross sections after cuts in
the jets$+\eslt$ ($0\ell$) channel are 1.9, 3.3 and 2.1 fb repectively for the
mSUGRA, CSB and RNS cases while SM BG lies at 1.2 fb. In Fig. \ref{fig:L}, we
show the required value of LHC13 integrated luminosity which is needed to
establish a $5\sigma$ signal, where in addition we also require at least 10
total signal events. From this plot, we see that just 8.3, 3.1 or 6.9 fb$^{-1}$
of integrated luminosity $L$ is needed to establish a first signal for the
mSUGRA, CSB and RNS benchmark models with $m_{\tg}=1.4$ TeV. The CSB benchmark
model has a somewhat larger signal cross section and hence requires somewhat
lower $L$ in the $0\ell$ channel as compared to the mSUGRA and RNS models since
its decay modes include more hadronic and fewer leptonic cascades.

In Fig. \ref{fig:sigma}, we also see the cross section after cuts for the
$1\ell$ channel. Even though one takes a leptonic branching fraction hit in this
channel, the numerous sources for a single additional isolated lepton lead to
cross sections after cuts which are comparable to those in the $0\ell$ channel.
For the $1\ell$ channel, RNS has the largest cross section $1.0$ fb while mSUGRA
and CSB are at the 0.8 fb level. This $1\ell+jets +\eslt$ channel will confirm
the signal which is already established in the $0\ell$ channel with just a few
additional (10-14) fb$^{-1}$ of integrated luminosity.

In Fig. \ref{fig:sigma} and \ref{fig:L}, we also show the cross section after
cuts and the required integrated luminosity for a $5\sigma$ signal for the OSSF,
SS, $3\ell$ and $4\ell$ channels. These multi-lepton channels all exhibit a
greater suppression due to multiple leptonic branching fractions as compared to
the $0\ell$ channel. For the $3\ell$ channel, background events come from
isolated leptons in the $b$-quark decays. With the requirement of at least 3
$b$-jets, we do not observe any events in our $t\bar{t}$ background for the
$4\ell$ channel. Also, in the multi-lepton channels, we see that the RNS model
yields the largest cross sections due to the large gluino branching fractions
into tops followed by $t\to bW$ and $W\to \ell\nu$ decay. From Fig. \ref{fig:L},
we see that typically $\sim 100$ fb$^{-1}$ is necessary to establish a signal in
the dilepton and trilepton channels while $\sim 10^3-10^4$ fb$^{-1}$ would be
required for a $5\sigma$ signal in the $4l$ channel.

\section{Establishing the LSP archetype}
\label{sec:LSP}

\subsection{Charged SUSY breaking}

\begin{figure}[!htb]
\begin{center}
\includegraphics[width=110mm]{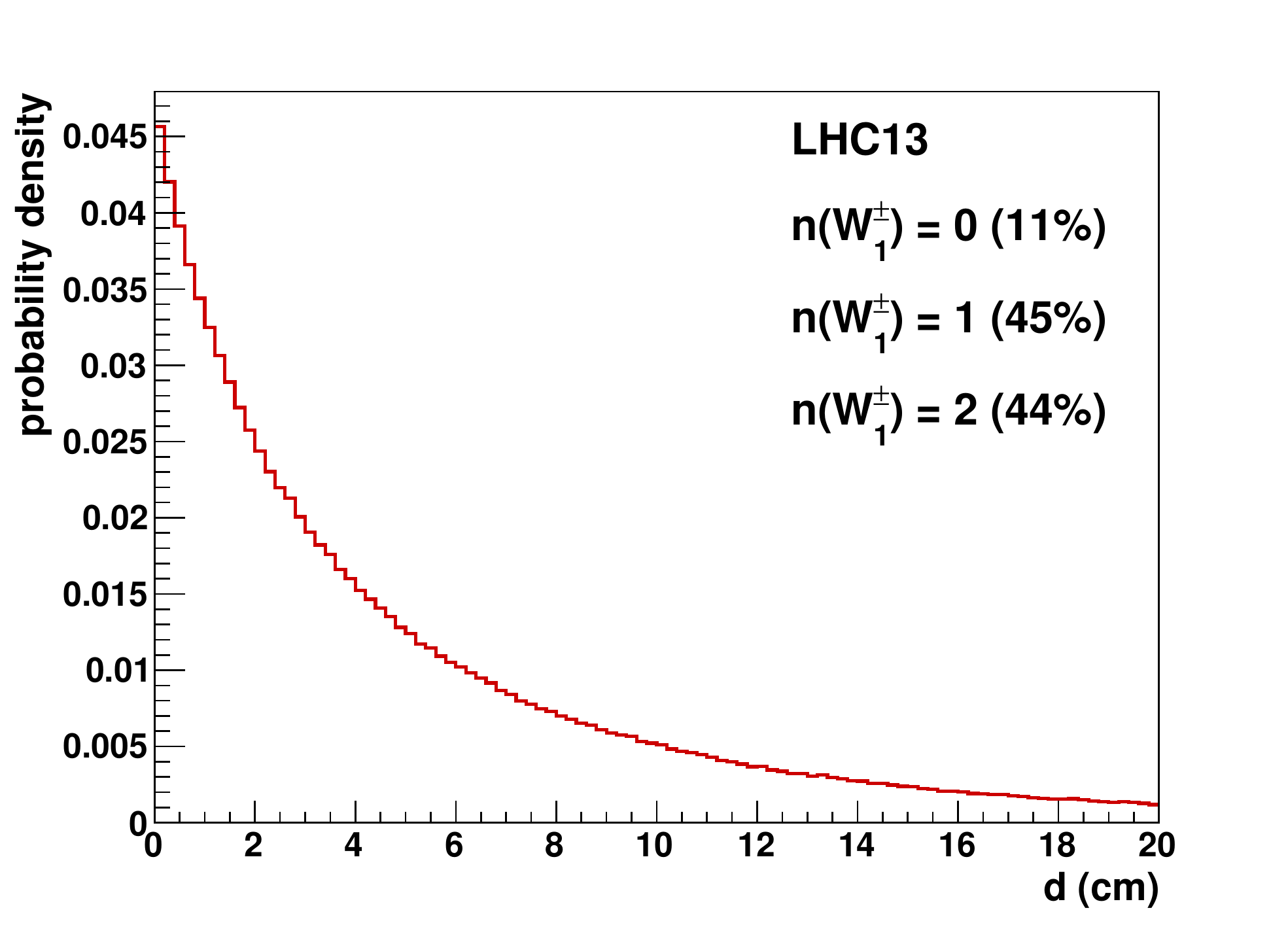}
\caption{Distance to the decay point from the interaction vertex for the CSB
benchmark model.}
\label{fig:chargino_displacement}
\end{center}
\end{figure}

One of the features of the CSB model is that $\charg1$ and the LSP are almost
degenerate with a mass difference $\Delta m = m_{\charg1} - m_{\neut1} \gtrsim
m_{\pi^{\pm}}$. $\charg1$ decays into charged pions at almost 100\% rate. The
reduced phase space also makes the chargino long-lived so that, once produced at
the interaction vertex, it travels a visible distance before it decays to soft
pions plus the LSP. Since the chargino is so massive, its velocity is borderline
relativistic leading to a highly-ionizing trail or track (HIT). The chargino
lifetime $\tau_{\tw_1}$ is extracted from Isajet and the actual lifetime of each
chargino is generated from the exponential decay law $e^{-t/ \tau_{\tw_1}}$.
Then the track length is computed from $d=\beta\gamma t$.

In Figure \ref{fig:chargino_displacement}, we display the histogram of the
distance travelled from the interaction vertex to the decay point of each
chargino. Here, we see that the typical length of each HIT is of order 2-20 cm.
We also display the percentage of events containing 0-2 charginos. We see that
90\% of the events passing our cuts contain either one or two charginos in each
event. The presence of one-or-more HITs in candidate SUSY events would be the
smoking gun signature of SUSY models with a wino-like LSP.

\subsection{Radiatively-driven naturalness (RNS)}

In the RNS benchmark model, it is emphasized\cite{bbh,ltr,rns@lhc} that the mass
gap between the $\tz_2$ and $\tz_1$ neutralinos is typically small: $\sim 10-30$
GeV which gives the inter-higgsino splitting. For our benchmark case, the value
is $ \Delta m = m_{\neut2} - m_{\neut1} = 27.4$ GeV. Notice this mass gap never
gets much below about 10 GeV since naturalness also provides upper bounds to the
gaugino masses via loop effects so that the higgsino-gaugino mass gap cannot
become arbitrarily large. The modest $\tz_2-\tz_1$ mass gap has important
consequences for phenomenology. It means that the $\tz_2$ always decays via
$3$-body modes $\tz_2\to\tz_1 f\bar{f}$ which is dominated by $Z^*$ exchange.
The decay mode $\tz_2\to\tz_1\ell^+\ell^-$ occurs at 3-4\% per lepton species,
but the OSSF dilepton pair which emerges from this decay always has invariant
mass kinematically bounded by $m_{\tz_2}-m_{\tz_1}$. This mass edge should be
apparent in gluino pair cascade decay events which contain an OSSF dilepton
pair.
\begin{figure}[!htb]
\begin{center}
\includegraphics[width=110mm]{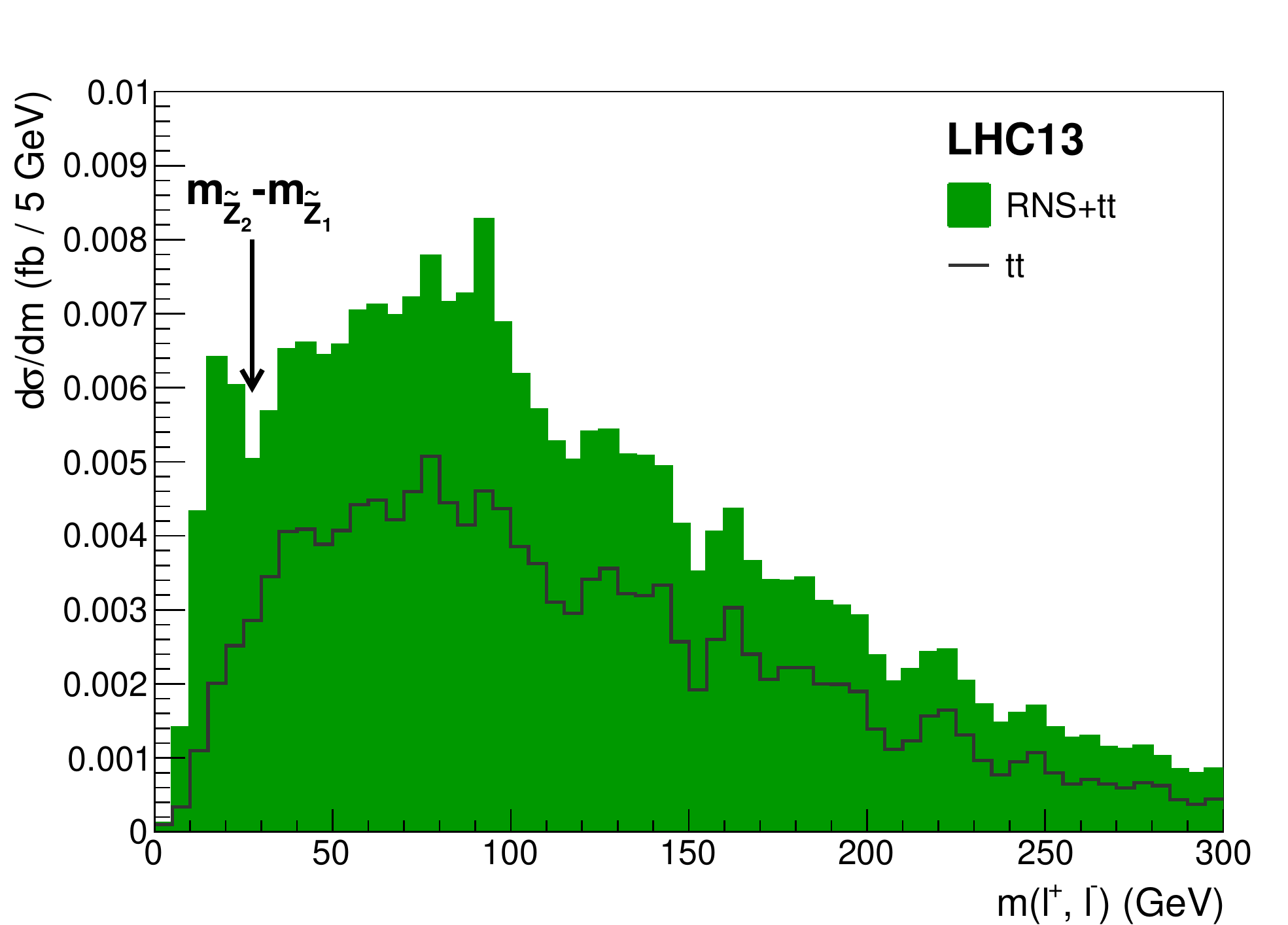}
\caption{Invariant mass of OSSF leptons. The dilepton mass edge and the $Z$ peak
are visible for the RNS model. We require $n(\textit{b-jets}) \geq 3$.}
\label{fig:invmassRNS}
\end{center}
\end{figure}

In Fig. \ref{fig:invmassRNS}, we show the invariant mass distribution of OSSF
dilepton pairs in gluino pair cascade decay events where we require the above
cuts but with $\eslt > \max(100\ {\rm GeV},0.2M_{eff})$ and $A_T>1200$ GeV and
the presence of an isolated OSSF dilepton pair. The black histogram shows the
expected continuum background distribution arising mainly from $t\bar{t}$
production while the green histogram shows signal plus BG for the RNS benchmark
model. The RNS signal is characterized by the distinct mass bump and edge below
about 30 GeV. This feature provides the smoking gun signature for SUSY models
with light higgsinos\cite{rns@lhc}. One can also see a peak at $m(\ell^+\ell^-
)\sim m_Z$ which arises from $\tw_2$ and $\tz_{3,4}$ two-body decays to a real
$Z$. The area under the $m(\ell^+\ell^-) < 30$ GeV portion is $\sim 0.025$ fb so
that of order 400 fb$^{-1}$ of integrated luminosity will be required before
this feature begins to take shape in real data.

\begin{figure}[!htb]
\begin{center}
\includegraphics[width=110mm]{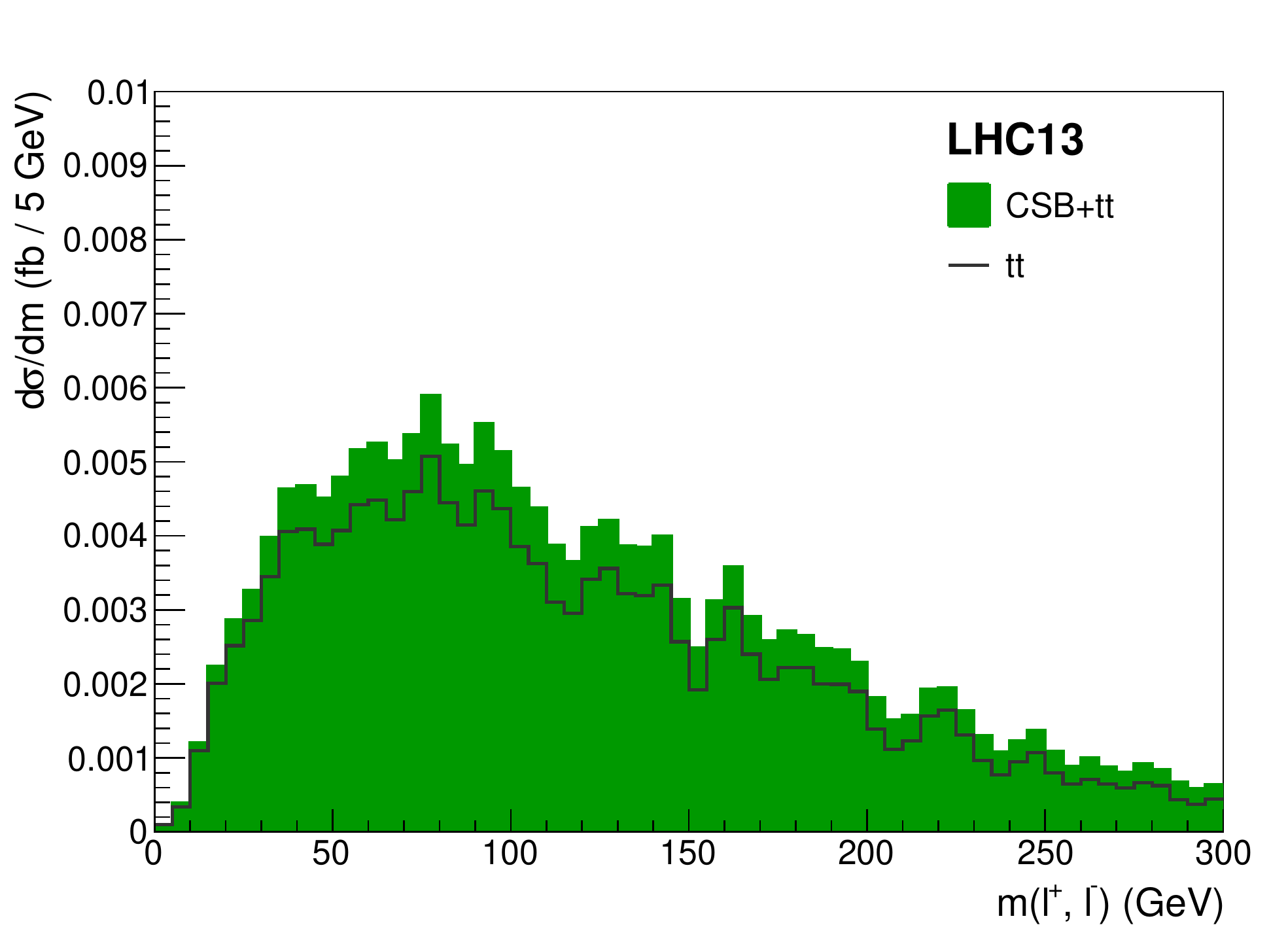}
\caption{Invariant mass of OSSF dileptons for the CSB model. 
We require $n(\textit{b-jets}) \geq 3$.}
\label{fig:invmassCSB}
\end{center}
\end{figure}

For comparison, in Fig. \ref{fig:invmassCSB} we show the same $m(\ell^+\ell^- )$
distribution for the case of the CSB benchmark. In the CSB case, first of all
there are far fewer $\ell^+\ell^-$ pairs present above background, and second
there is no obvious structure to the signal distribution: we expect just a
continuum.

The second smoking gun signature for models with a higgsino LSP is the presence
of same-sign diboson (SSdB) events which are from wino pair
production\cite{lhcltr,rns@lhc}. In this case, the production reaction is
typically $pp\to\tw_2\tz_4$ followed by $\tw_2\to\tw_1\tz_{1,2}$ and
$\tz_4\to\tw_1^\pm W^\mp$. The Majorana nature of the $\tz_4$ leads to equal
amounts of same-sign and opposite sign dilepton events. Note that these SSdB
events contain minimal jet activity-- only that arising from initial state QCD
radiation-- as opposed to SS dilepton events from gluino and squark cascade
decays which should be rich in the presence of additional high $p_T$ jets.

\subsection{mSUGRA/CMSSM}

\begin{figure}[!htb]
\begin{center}
\includegraphics[width=110mm]{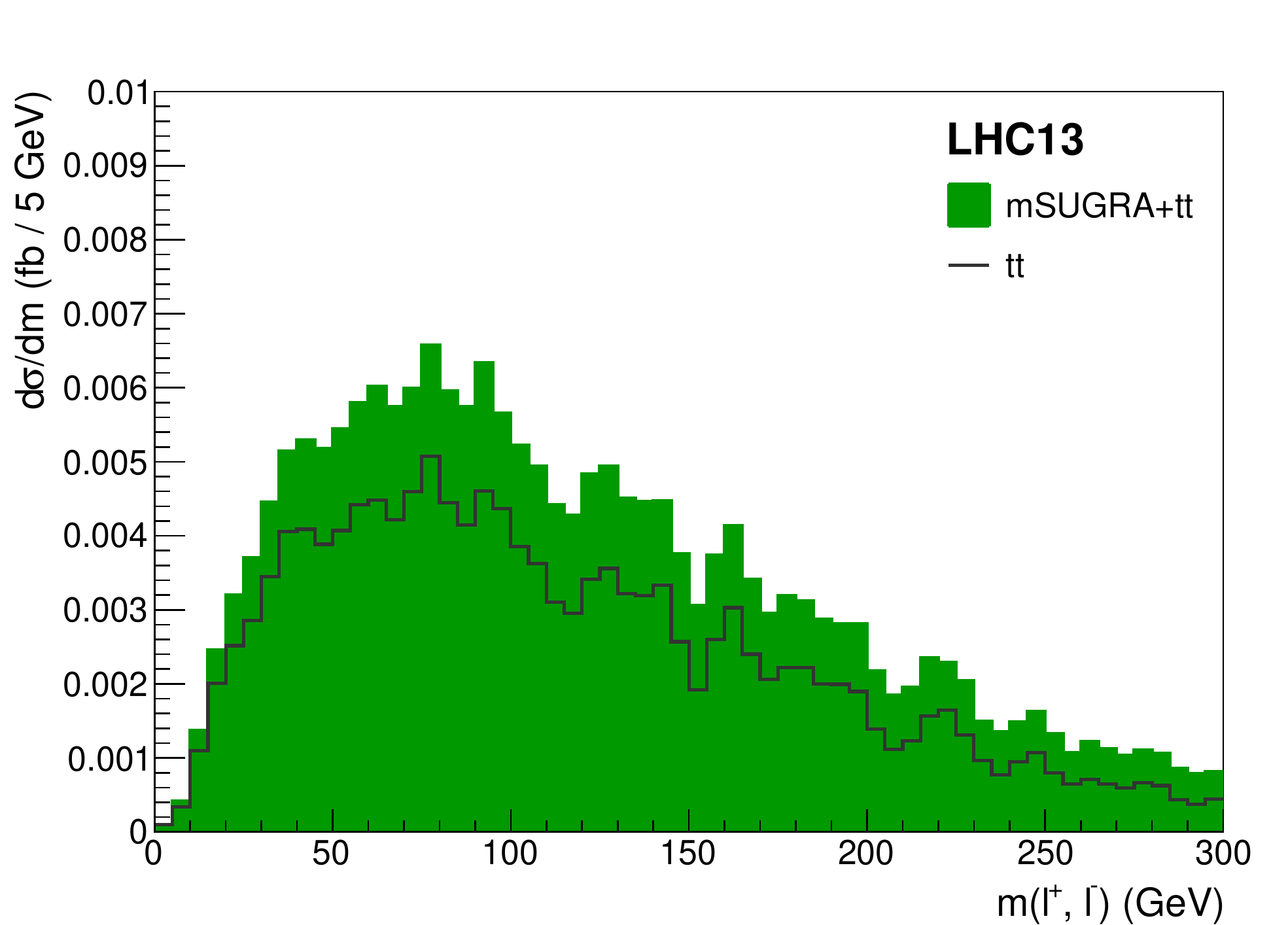}
\caption{Invariant mass of OSSF dileptons for the mSUGRA model. 
We require $n(\textit{b-jets}) \geq 3$.}
\label{fig:invmassSUGRA}
\end{center}
\end{figure}

For the mSUGRA/CMSSM benchmark model with a 1.4 TeV gluino, then we expect the
production of the usual multi-lepton+multi-jet + $\eslt$ cascade decay signatures
as shown in Fig. \ref{fig:sigma}. For the case of the mSUGRA benchmark, the mass
gap between the wino-like $\tz_2$ and the bino-like $\tz_1$ is 225.5 GeV so that
the $\tz_2\to\tz_1 h$ (spoiler) decay mode is open. This two-body decay
dominates the $\tz_2$ branching fraction and so we expect no additional
structure in the dilepton invariant mass distribution. The $m(\ell^+\ell^- )$
distribution for the mSUGRA benchmark point is shown in Fig.
\ref{fig:invmassSUGRA}. While no characteristic dilepton structure is apparent,
it may be possible instead to pull out the presence of $h\to b\bar{b}$ decays in
the mSUGRA cascade decay events where $m(b\bar{b})\sim m_h$\cite{oldhbb}. 

\section{Conclusions}
\label{sec:conclude}

During run 1 of the LHC at $\sqrt{s}=7-8$ TeV, the Standard Model was vigorously
confirmed in both the electroweak and QCD sectors and the Higgs boson was
discovered at $m_h\simeq 125$ GeV. The presence of a bonafide fundamental scalar
particle cries out for a mass stabilization mechanism of which the simplest and
most elegant one is supersymmetry. Unfortunately, no SUSY particles have yet
appeared leading to mass limits for the gluino particle of $m_{\tg}\agt 1.3$
TeV.

LHC Run 2 with $\sqrt{s}=13$ TeV has begun! New vistas in SUSY parameter space
are open for exploration. While naturalness allows for gluinos as high as 4-5
TeV (with $\Delta_{EW}<30$), it is yet true that naturalness (mildly via higher
order contributions) prefers gluinos as light as possible. Motivated by these
circumstances, we considered how SUSY discovery would unfold in three SUSY
archetype models with a bino-, wino- and higgsino-like LSP each with a 1.4 TeV
gluino, just beyond present bounds.

We find that SUSY discovery could already arise at the $5\sigma$ level with just
$3-8$ fb$^{-1}$ of integrated luminosity via the $\ge 3\ b-{\rm jet}+\eslt$
channel. Confirmation would soon follow in the $\ge 3\ b-{\rm jet}+1-\ell+\eslt$
channel. Further confirmation in the 2-3 lepton channels will require $\sim 100$
fb$^{-1}$. The CSB benchmark case would immediately be identified by the
presence of one or more highly ionizing tracks in each signal event due to
long-lived wino-like charginos which undergo delayed decays to a wino-like LSP.
No such HITs should be apparent in signal events from the mSUGRA or RNS
archetype models. Instead, the RNS archtype would be signalled by a gradual
buildup of structure in the $m(\ell^+\ell^- )$ OSSF dilepton mass distribution,
where the $m(\ell^+\ell^- )<m_{\tz_2}-m_{\tz_1}$ mass edge along with a $Z$ peak
should be apparent with $\sim 100-1000$ fb$^{-1}$. In the RNS case, the gluino
cascade decay events should ultimately be accompanied by the presence of
same-sign diboson events arising from wino pair production. 

For the mSUGRA archtype with a bino-like LSP, then we expect the usual
assortment of gluino-pair-initiated cascade decay multilepton+jets + $\eslt$
events but without HITs and without any apparent structure in the
$m(\ell^+\ell^- )$ distribution. However, the presence of Higgs bosons lurking
within the cascade decay events may be a distinguishing feature. On to data from
LHC13!

\acknowledgments 

The authors would like to thank CETUP* (Center for Theoretical Underground
Physics and Related Areas), for its hospitality and partial support during the
2015 Summer Program. 
This work was supported in part by the US Department of Energy, Office of
High Energy Physics.

\end{document}